# Anomaly Detections in Internet traffic Using Empirical Measures

A.S.Syed Navaz, S.Gopalakrishnan, R.Meena

*Abstract—* Introducing Internet traffic anomaly detection mechanism based on large deviations results for empirical measures. Using past traffic traces we characterize network traffic during various time-of-day intervals, assuming that it is anomaly-free. Throughout, we compare the two approaches presenting their advantages and disadvantages to identify and classify temporal network anomalies. We also demonstrate how our framework can be used to monitor traffic from multiple network elements in order to identify both spatial and temporal anomalies. We validate our techniques by analyzing real traffic traces with time-stamped anomalies.

## I. INTRODUCTION

### A. Spatio-temporal network

The need for spatio-temporal network arises naturally when dealing with problems such as speech recognition and time series prediction where the input signal has an explicit temporal aspect. We have demonstrated that certain tasks that do not have an explicit temporal aspect can also be processed advantageously with neural networks capable of dealing with temporal information. In particular, we have proposed that converting static patterns into time-varying (spatio-temporal) signals by scanning the image would lead to a number of significant advantages.

- An obvious but significant advantage of scanning approach is that it naturally leads to a recognition system that is shift invariant along the temporalized axis(es).
- The spatio-temporal approach explicates the image geometry since the local spatial relationships in the image along the temporalized dimension are naturally expressed as local temporal variations in the scanned input.
- In the spatio-temporal approach a spatial dimension is replaced by the temporal dimension and this leads to models that are architecturally less complex than similar models that use two spatial dimensions. This reduction of complexity occurs because the spatial extent of any feature in an object's image is much less than the spatial extent of the object's image. If we assume that internal nodes act as feature detectors then the moving receptive field of an internal node in the spatio-temporal model leads to an effective tessellation of the feature detector over the image without the actual (physical) replication of the feature detector.
- Work on visual pattern recognition often treats an image as a static fixed length two-dimensional pattern. This is unrealistic since in general an agent must scan its environment to locate and identify objects of interest. Observe that even reading text involves processing a continuous stream of visual data having an essentially arbitrary extent. The scanning approach allows a visual pattern recognition system to deal with inputs of arbitrary extent.

The effectiveness of the above ideas has been demonstrated in the dissertation work of thomas fontaine who designed and trained a system for recognizing sequences of handwritten digits. The system has a 96% recognition rate on a dataset of 2,700 isolated digits provided by usps and a 96.5% recognition rate on a set of 207,000 isolated digits provided by nist. On a set of 540 real-word zip code images provided by usps, the system achieved a raw accuracy of 66.0%

**Spatio temporal gene expression** is the activation of genes within specific tissues of an organism at specific times during development. Gene activation patterns vary widely in complexity. Some are straightforward and static, such as the pattern of *tubulin,* which is expressed in all cells at all times in life. Some, on the other hand, are extraordinarily intricate and difficult to predict and model, with expression fluctuating wildly from minute to minute or from cell to cell. Spatiotemporal variation plays a key role in generating the diversity of cell types found in developed organisms; since the identity of a cell is specified by the collection of genes actively expressed within that cell, if gene expression was uniform spatially and temporally, there could be at most one kind of cell.

### B. Anomaly detection

In this work we focus on anomaly detection and in particular on *statistical anomaly detection*, where statistical methods are used to assess deviations from normal operation. Our main contribution is the introduction of a new statistical traffic anomaly detection framework that relies on identifying deviations of the empirical measure of some underlying stochastic process characterizing system behavior. In contrast with other approaches , we are not trying to characterize the abnormal operation, mainly because it is too complex to identify all the possible anomalous instances (especially those that have never been observed). Instead we observe past system behavior and, assuming that it is anomaly-free, we obtain a statistical characterization of "normal behavior." Then, using this knowledge we continuously monitor the system to identify time instances where system behavior does not appear to be

Manuscript received on February, 2013.
A.S.Syed Navaz, Asst.Professor, Department of Computer Applications, Periyar University/ Muthayammal College of Arts & Science/ Namakkal, India.
S.Gopalakrishnan, Asst.Professor, Department of Computer Science, Periyar University/ Muthayammal College of Arts & Science/ Namakkal, India.
R.Meena, Asst.Professor, Department of Computer Applications, Periyar University/ Muthayammal College of Arts & Science/ Namakkal, India.

normal. The novelty of our approach is in the way we characterize normal behavior and in how we assess deviations from it. More specifically, we propose two methods to characterize normal behavior:

- ✓ A *model-free* approach employing the method of types to characterize the type (i.e., empirical measure) of an independent and identically-distributed (i.i.d.) Sequence of appropriately averaged system activity
- ✓ A *model-based* approach where system activity is modeled using a *markov modulated process (mmp)*. Given these characterizations, we employ the theory of *large deviations (ld)* and decision theory results to assess whether current system behavior *deviates* from normal.

ld theory provides a powerful way of handling rare events and their associated probabilities with an asymptotically exact exponential approximation. The key technical results we rely upon are sanov's theorem in the model-free approach, a related result for the empirical measure of a markov process for the model-based case, and hoeffding's composite hypothesis testing rule for assessing deviations from normal activity.

The proposed framework is general enough to also take into account spatial information. By combining observations at different locations of a network we are able to construct multi-dimensional stochastic processes to characterize system behavior. Both our model-free and our mmp approaches can work with vector characterizations and identify spatio-temporal anomalies (as both temporal and spatial information is preserved).

The methods we present are statistical; as a result, our approach has the potential of detecting novel anomalies, such as previously unseen attacks. This is crucial for network security as new types of attacks are constantly being engineered. A novel feature of our approach is that it compares subtle distributional differences between the reference traffic characterization and Observed traffic traces. As we will see, this is critical as it enables us to detect attacks—including some short-lived ones or anomalies that traverse different locations of a network—that do not result in significant changes in traffic volume. First or second moments of traffic measurements would be too insensitive to these types of attacks. A distinctive feature of our approach is that it appears able to identify temporal and spatial anomalies in short time-scales as opposed to techniques working over much longer time-scales or others that try to identify spatial anomalies collapsing the temporal correlation of network feature samples. As we will also describe later, the only infrastructure requirement in order to deploy our method is a simple counter

### C. Empirical Measures for Anomaly Detection

As was mentioned before, the size of the alphabet and the number of states of the MMP for the Abilene data set is small when only temporal information is considered. Thus, it is easy to monitor subnets of PoPs (of low dimensionality) by specifying the group of PoPs of interest and the role of each PoP (origin or destination). We present results for two case studies with different spatial characteristics. We apply our framework to: (a) flows that originate (end) from (at) PoPs that are 1-hop neighbors and (b) flows that originate (end) from (at) PoPs that are many hops away from each other. In the first case study, the flows originate (end) at the Sunny Valley (SNVA) PoP with destination (originating from) the PoPs in its vicinity. We illustrate instances of the identification of anomalies applying the model-free and the model based methods, respectively. The values of the parameters for the two methods are obtained from the temporal anomaly detection examples. Table II reports the detection and false alarm rates we achieved. It is worth noticing that the detection rate reached 100% and the false alarms rate was very low (lower than the values when only temporal anomalies were studied). This is due to two main reasons: (a) instantaneous high values in the time-series of observations that do not necessarily indicate attacks are smoothed due to time averaging, and (b) attacks may have temporal and/or spatial correlation.

### D. Congestion Traffic Minimization

We provided two different approaches, a model-free and a model-based one. The model-free method works on a longer time-scale processing traces of traffic aggregates over a small time interval. Using an anomaly-free trace it derives an associated probability law. Then it processes current traffic and quantifies whether it conforms to this probability law. The model-based method constructs a Markov modulated model of anomaly-free traffic measurements and relies on large deviations asymptotics and decision theory results to compare this model to ongoing traffic activity. We presented a rigorous framework to identify traffic anomalies providing asymptotic thresholds for anomaly detection. In our experimental results the model-free approach showed a somewhat better performance than the model-based one. This may be due to the fact that the former gains from the aggregation over a time-bucket in addition to the fact that the latter one requires the estimation of more parameters, hence, it may introduce a larger modeling error. For future work, it would be interesting to analyze the robustness of the anomaly detection mechanism to various model parameters.

Since we monitor the detailed distributional characteristics of traffic and do not rely on the mean or the first few moments we are confident that our approach can be successful against new types of (emerging) temporal and spatial anomalies.

Our method is of low implementation complexity (only an additional counter is required), and is based on first principles, so it would be interesting to investigate how it can be embedded on routers or other network devices.

## II. SYSTEM ANALYSIS

### A. Existing System

Although significant progress has been made in network monitoring instrumentation, automated on-line traffic anomaly detection is still a missing component of modern network security and traffic engineering mechanisms. Network anomaly detection approaches can be broadly grouped into two classes: signature-based anomaly detection where known patterns of past anomalies are used to identify ongoing anomalies and anomaly detection which identifies patterns that substantially deviate from normal patterns of operation. Earlier work has showed that systems based on pattern matching had detection rates below 70% Furthermore, such systems need constant (and expensive)

### B. Proposed System

In this work we focus on anomaly detection and in particular on statistical anomaly detection, where statistical methods are used to assess deviations from normal operation. Our main contribution is the introduction of a new statistical traffic anomaly detection framework that relies on identifying deviations of the empirical measure of some underlying stochastic process characterizing system behavior.

In contrast with other approaches we are not trying to characterize the abnormal operation, mainly because it is too complex to identify all the possible anomalous instances (especially those that have never been observed). Instead we observe past system behavior and, assuming that it is anomaly free, we obtain a statistical characterization of "normal behavior." Then, using this knowledge we continuously monitor the system to identify time instances where system behavior does not appear to be normal. The novelty of our approach is in the way we characterize normal behavior and in how we assess deviations from it. More specifically, we propose two methods to characterize normal behavior:

- A model-free approach employing the method of types to characterize the type (i.e., empirical measure) of an independent and identically distributed (i.i.d.) sequence of appropriately averaged system activity, and
- A model-based approach where system activity is modeled using a Markov Modulated Process (MMP)

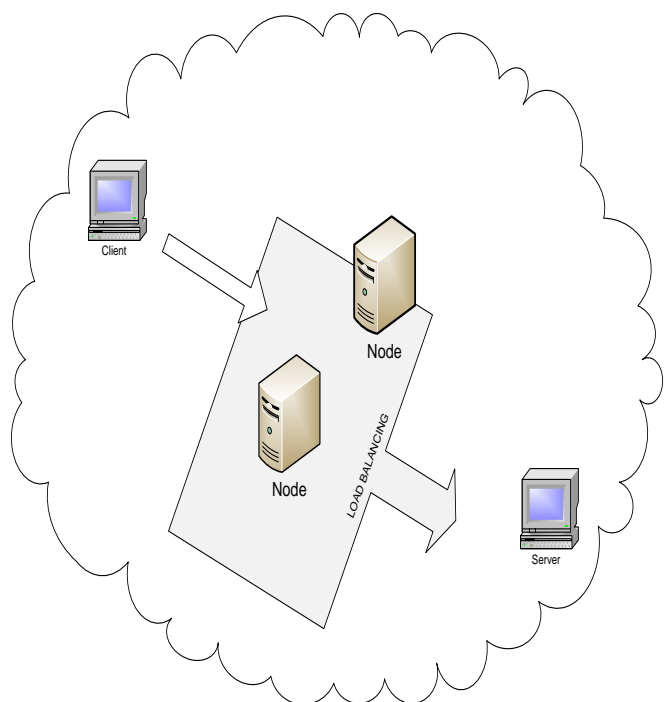

Fig 1. Client to Server Load Balancing Architecture Diagram

### III. SYSTEM DESIGN

The system design involves input design and output design and modular of proposed system.
- Network Model
- Model-free approach
- Model Based Markov modulated process
- Graphical process

*A. Network Model*

➢ Generally, the channel quality is time-varying. For the ser-AP association decision, a user performs multiple samplings of the channel quality

➢ only the signal attenuation that results from long-term channel condition changes are utilized our load model can accommodate various additive load definitions such as the number of users associated with an AP.
➢ It can also deal with the multiplicative user load contributions.

*Client Model :*

➢ A **client** is an application or system that accesses a remote service on another computer system, known as a server, by way of a network.
➢ The term was first applied to devices that were not capable of running their own stand-alone programs, but could interact with remote computers via a network.
➢ These dumb terminals were clients of the time-sharing mainframe computer

*Server model:*

➢ In computing, a **server** is any combination of hardware or software designed to provide services to clients.
➢ When used alone, the term typically refers to a computer which may be running a server operating system, but is commonly used to refer to any software or dedicated hardware capable of providing services.

*B. Model-free approach*

➢ we discuss our model-free approach and provide the structure of an algorithm to detect temporal network anomalies.
➢ As noted in the Introduction we focus on traffic at points of interest in the network, even though our approach is general enough to be applied to any trace of system activity.
➢ We assume that the *traffic trace* we monitor (in bits/bytes/packets/ flows per time unit), corresponding to a specific time-of-day interval, can be characterized by a stationary model over a certain period (e.g., a month) if no technological changes (e.g., link bandwidth upgrades) have taken place.
➢ Consider a time series of $x_1$................$x_n$ traffic activity (say, in bits/bytes/packets/flows per sample).

*C. Model Based Markov modulated process*

➢ The approach of aggregated traffic over a time bucket to yield an i.i.d. sequence.
➢ One potential disadvantage of this aggregation is that it increases the response time to an anomaly since data is being processed on the slower time-scale of time buckets.
➢ In this section, the question we are seeking to answer is whether it is possible to process data on the timescale we collect them.
➢ To that end, and because the i.i.d. assumption will no longer hold, we will impose some more structure on the stochastic nature of the traffic time-series.
➢ In particular, we will assume a Markovian structure as it is tractable and has been shown to represent traffic well, at least for the purpose of estimating distribution-dependent metrics like loss probabilitie

*An MMP Model :*

➢ We start again with a time series of traffic ac-tivity $1; \cdots; n$ during a small time interval (several hours) which we will model as an MMP process. Such a process is characterized by an underlying Markov chain with transition probability matrix.

➢ MMPs, when the state is "hidden", are also known in the literature as hidden Markov models (HMMs). We restrict ourselves to models in which the ranges of possible observations corresponding to different states are disjoint.

*D. Graphical process*

➢ We represent the anomaly detection over the IP that will show on graph.
➢ Graph process of MMP model.
➢ To reduced the traffic over the network.
➢ And reduced the time latency

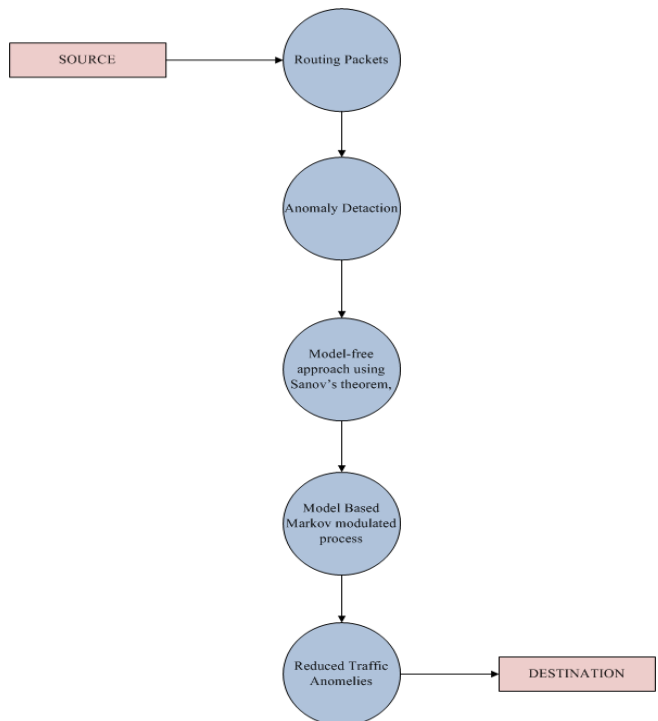

Fig 2. Anomoly Detection

## IV. CONCLUSION

As introduced a general distributional fault detection scheme able to identify a large spectrum of temporal anomalies from attacks and intrusions to various volume anomalies and problems in network resource availability. We then showed how this framework can be extended to incorporate spatial information, resulting in robust spatio-temporal anomaly detection in large scale operational networks. Although most of the proposed anomaly detection frameworks are able to identify temporal or spatial anomalies we are able to identify both as we preserve both the temporal and spatial correlation of network feature samples.

As provided two different approaches, a model-free and a model-based one. The model-free method works on a longer time-scale processing traces of traffic aggregates over a small time interval. Using an anomaly-free trace it derives an associated probability law. Then it processes current traffic and quantifies whether it conforms to this probability law. The model-based method constructs a Markov modulated modeof anomaly-free traffic measurements and relies on large deviations asymptotics and decision theory results to compare this model to ongoing traffic activity.

As presented a rigorous framework to identify traffic anomalies providing asymptotic thresholds for anomaly detection. In our experimental results the model-free approach showed a somewhat better performance than the model-based one. This may be due to the fact that the former gains from the aggregation over a time-bucket in addition to the fact that the latter one requires the estimation of more parameters, hence, it may introduce a larger modeling error.

## REFERENCES

[1] M. Roesch, "Snort—Lightweight intrusion detection for networks," in LISA '99: Proc. 13th USENIX Conf. System Administration, Seattle, WA, Nov. 1999, pp. 229–238.
[2] Paxson, "Bro: A system for detecting network intruders in real time," Computer Networks, vol. 31, no. 23–24, pp. 2435–2463, 1999.
[3] P. Barford, J. Kline, D. Plonka, and A. Ron, "A signal analysis of network traffic anomalies," in Proc. ACM SIGCOMM Workshop on Internet Measurement, Marseille, France, Nov. 2002, pp. 71–82.
[4] R. Lippmann, D. Fried, I. Graf, J. Haines, K. Kendall, D. McClung, D. Weber, S. Webster, D.Wyschogrod, R. Cunningham, and M. Zissman, "Evaluating intrusion detection systems: The 1998 DARPA off-line intrusion detection evaluation," in Proc. DARPA Information Survivability Conf. and Expo., Los Alamitos, CA, Jan. 2000, pp. 12–26.
[5] R. Lippmann, J. W. Haines, D. J. Fried, J. Korba, and K. Das, "The 1999 DARPA off-line intrusion detection evaluation," Computer Networks, vol. 34, no. 4, pp. 579–595, 2000.
[6] Yegneswaran, J. T. Giffin, P. Barford, and S. Jha, "An architecture for generating semantics-aware signatures," in USENIX Security Symp., Baltimore, MD, Jul. 2005, pp. 97–112.
[7] I.Dembo and O. Zeitouni, Large Deviations Techniques and Applications, 2nd ed. New York: Springer-Verlag, 1998.
[8] W. Hoeffding, "Asymptotically optimal tests for multinomial distributions," Ann. Math. Statist. vol. 36, pp. 369–401, 1965.
[9] I.Paschalidis and S. Vassilaras, "On the estimation of buffer overflow probabilities from measurements," IEEE Trans. Inf. Theory, vol. 47, no. 1, pp. 178–191, 2001.
[10] I.Paschalidis and S. Vassilaras, "Model-based estimation of buffer overflow probabilities from measurements," in Proc. ACM SIGMETRICS 2001/Performance 2001 Conf., Cambridge, MA, Jun. 16–20, 2001, pp. 154–163.

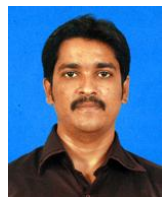

**A.S.Syed Navaz** received BBA from Annamalai University, Chidambaram 2006, M.Sc Information Technology from KSR College of Technology, Anna University Coimbatore 2009, M.Phil in Computer Science from Prist University, Thanjavur 2010 and M.C.A from Periyar University, Salem 2010 .Currently he is working as an Asst.Professor in Department of Computer Applications, Muthayammal College of Arts & Science, Namakkal. His area of interests are Wireless Communications, Computer Networks and Mobile Communications.

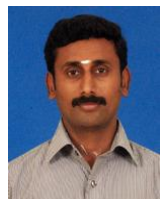

**S.Gopalakrishnan** received B.Sc Computer Science & M.Sc Computer Science from Nehru Memorial College, Bharathidasan University, Trichy 2002 & 2004, M.Phil in Computer Science from Periyar University, Salem 2009 & B.ed Computer Science from Rainbow College of Education, Namakkal 2011. Currently he is working as an Asst.Professor in Department of Computer Science, Muthayammal College of Arts & Science, Namakkal. His area of interests are Computer Networks and Mobile Communications.

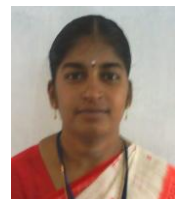

**R.Meena** received BCA from Bharathiyar Arts & Science College, Periyar University, Salem 2007, M.Sc Computer Science from Muthayammal College of Arts & Science, Periyar University, Salem 2009 & M.Phil in Computer Science from Thanthai Hans Rover College, Bharathidasan University, Perambalur Currently she is working as an Asst.Professor in Department of Computer Applications, Muthayammal College of Arts & Science, Namakkal. Her area of interests are Computer Networks and Mobile Communications.